\documentclass[sigconf]{acmart}

\copyrightyear{2027}
\acmYear{2027}
\setcopyright{cc}
\setcctype{by}
\acmConference[WSDM '27]{Proceedings of the Twentieth ACM International Conference on Web Search and Data Mining}{February 15--19, 2027}{Hong Kong, Hong Kong}
\acmBooktitle{Proceedings of the Twentieth ACM International Conference on Web Search and Data Mining (WSDM '27), February 15--19, 2027, Hong Kong, Hong Kong}
\acmDOI{10.1145/3862322.3863432}
\acmISBN{979-8-4007-3117-4/2027/02}

\usepackage{graphicx}
\usepackage{subcaption} 
\usepackage{caption}
\usepackage{float}
\usepackage{multirow}
\usepackage{algorithm}
\usepackage{algorithmic}
\usepackage{comment}

\begin{document}

\title[GateSID: Balancing Semantic and Collaborative Signals]%
{GateSID: One Gate to Balance Semantic and Collaborative Signals across the Item Lifecycle}

\author{Hai Zhu}
\authornote{Corresponding author.}
\email{zhuhai@alibaba-inc.com}
\affiliation{%
  \institution{Alibaba International Digital Commercial Group}
  \city{Hangzhou}
  \country{China}
}

\author{Lei Shen}
\email{kenny.sl@alibaba-inc.com}
\affiliation{%
  \institution{Alibaba International Digital Commercial Group}
  \city{Hangzhou}
  \country{China}
}

\author{Bing Wang}
\email{lingfeng.wb@alibaba-inc.com}
\affiliation{%
  \institution{Alibaba International Digital Commercial Group}
  \city{Hangzhou}
  \country{China}
}

\author{Xiaoyi Zeng}
\email{yuanhan@alibaba-inc.com}
\affiliation{%
  \institution{Alibaba International Digital Commercial Group}
  \city{Hangzhou}
  \country{China}
}

\renewcommand{\shortauthors}{Hai Zhu et al.}

\begin{abstract}
In recommender systems, collaborative signals work well for popular items but are nearly absent for cold-start ones. Semantic content helps new items, yet over-reliance on it blurs the behavioral distinctions of popular items. Existing Semantic ID (SID) based methods handle this trade-off only partially, because their fusion of the two signals is rigid and item-agnostic. We present \textbf{GateSID}, a ranking framework that takes SID from an upstream tokenizer and adapts how each SID is used to the item's maturity through a single shared gate. The gate drives three mechanisms. Gating-Fused Shared Attention (GFSA) fuses the attention distributions of the two modalities rather than their representations. Gate-Regulated Contrastive Alignment (GRCA) strengthens semantic-behavioral alignment for items with weak collaborative evidence and relaxes it when collaborative structure is reliable. Gate-Regulated Semantic Resolution (GRSR) uses coarse shared SID prefixes for items with weak behavioral evidence and opens deeper SID levels when the gate indicates more reliable collaborative evidence. On a large-scale industrial dataset, GateSID leads the strongest SID-enhanced baseline by +0.27\% CTR AUC and +0.36\% CTCVR GAUC, with gains balanced across item groups rather than trading cold-start for popular-item performance; on the public Taobao-MM benchmark, it also leads overall and gives the strongest new-item results under both DIN and RankMixer backbones. In a one-month online A/B test, GMV improves by 2.6\%, CTR by 1.1\%, and Order by 1.6\%, at a cost of less than 5 ms of additional latency. We also provide a systematic analysis of SID design choices, covering embedding type, codebook configuration, and fusion strategy.
\end{abstract}


\maketitle

\section{Introduction}
\begin{figure*}[t]
    \centering
    \begin{subfigure}[t]{0.30\textwidth}
        \centering
        \includegraphics[width=\linewidth]{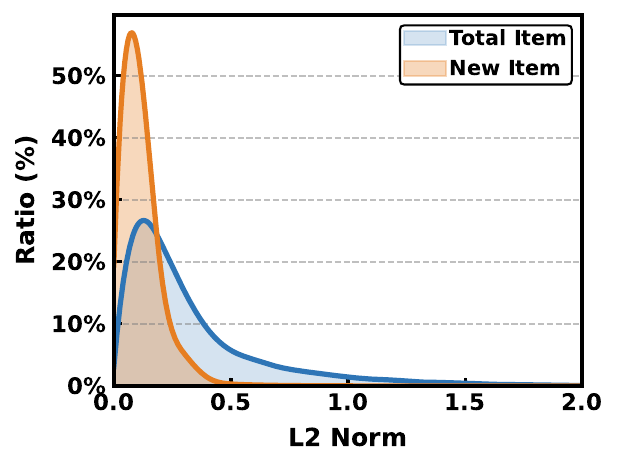}
        \caption{L2 norm distribution.}
        \label{fig:l2_norm}
    \end{subfigure}
    \hspace{0.025\textwidth}
    \begin{subfigure}[t]{0.30\textwidth}
        \centering
        \includegraphics[width=\linewidth]{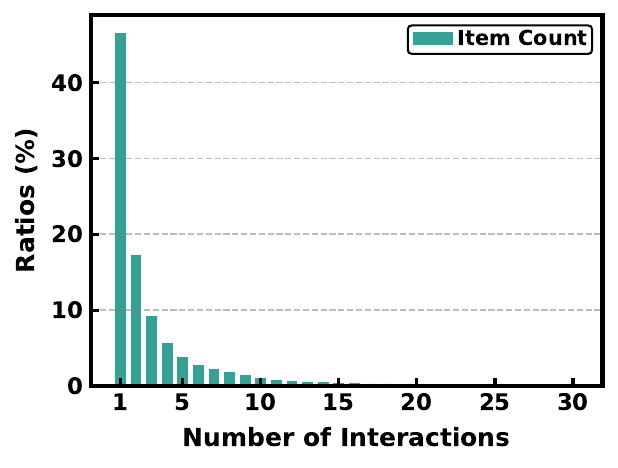}
        \caption{Interaction frequency distribution.}
        \label{fig:interaction}
    \end{subfigure}
    \hspace{0.025\textwidth}
    \begin{subfigure}[t]{0.30\textwidth}
        \centering
        \includegraphics[width=\linewidth]{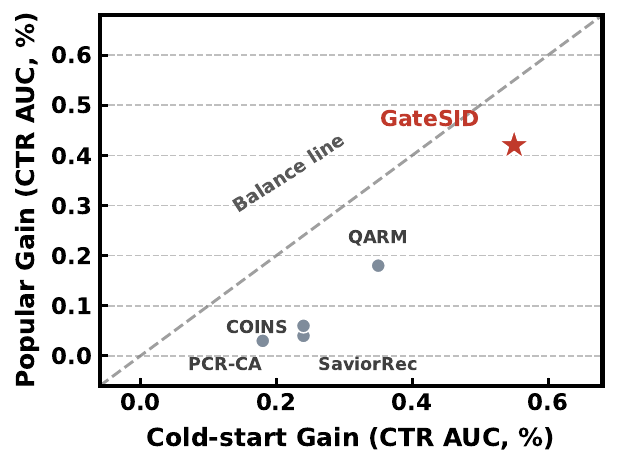}
        \caption{CTR AUC gains by item group.}
        \label{fig:tradeoff}
    \end{subfigure}
    \caption{Motivation analysis for lifecycle imbalance in item representations.}
    \label{fig:newitem_distribution}
\end{figure*}

Recommender systems serve items across very different maturity stages, from newly uploaded products with little feedback to popular items with abundant behavioral evidence \cite{zhu2025rankmixer,deng2025onerec,wang2026sort,zhu2025query,zhang2026onetrans}. This creates a trade-off between collaborative and semantic signals across the item lifecycle. Collaborative signals are highly discriminative when enough interactions have accumulated, but become unreliable for cold-start and long-tail items; semantic content provides priors for cold-start and long-tail items, while over-reliance on it can blur the behavioral distinctions learned from rich interactions. Figures~\ref{fig:l2_norm} and~\ref{fig:interaction} illustrate the cold-start and long-tail end of this lifecycle: 20\% of new items have zero-norm collaborative embeddings, and 46\% of the items in our data have only a single interaction record. This shifts the focus from simply injecting semantic information to estimating the appropriate degree of semantic intervention at each maturity stage.

Multimodal content, such as titles, images, and descriptions, offers a natural source of semantic priors \cite{li2025bbqrec,sheng2024enhancing,yan2025mim,bai2025chime,hu2025alphafuse,ganhor2024multimodal}.  Multimodal models \cite{wang2024qwen2} can encode such content, and Semantic ID (SID) methods \cite{zhao2025coins,tan2025pcr,yao2025saviorrec,liu2025best,zhang2025gpr} make it practical for ranking by discretizing multimodal embeddings into trainable code embeddings \cite{rajput2023recommender,deng2025onerec}. Compared with using fixed multimodal embeddings directly, SID provides a compact and trainable interface between content semantics and recommender systems. In this paper, we treat SID construction as an upstream module and focus on the ranking model after SID representations are available: \textbf{how to adaptively balance collaborative and semantic signals as items move from limited feedback to mature behavioral evidence}.

However, existing ranking methods built on SID often leave this lifecycle balance implicit. A lifecycle-aware ranking model can treat three choices as item specific: the semantic contribution to ranking, the alignment intensity between the semantic and collaborative spaces, and the granularity at which SID is read. Existing methods instead apply one fixed recipe to every item, so they benefit from semantics on new and long-tail items but have limited control over semantic interference on mature items.

Motivated by this gap, we introduce GateSID, a framework that adapts the weight of semantic signals to each item's maturity inside the ranking model. GateSID is plug and play for ranking models that already use SID representations. At the core of GateSID is a shared gate, computed from the item embedding and lifecycle statistics. This gate coordinates three modules: GFSA mixes SID and item attention so semantic cues enter sequence modeling to the needed extent; GRCA adjusts the alignment strength between SID and item embeddings; and GRSR selects coarse or fine SID granularity according to item maturity. Because the three modules read the same gate, GateSID adds only a small gating network to the ranking model and keeps one consistent notion of item maturity across attention, alignment, and granularity, rather than three separately tuned heuristics. Figure~\ref{fig:tradeoff} summarizes the outcome: existing methods show gains skewed toward cold-start items, whereas GateSID improves both groups, by +0.55\% on cold-start and +0.42\% on popular items. We summarize the main contributions of GateSID as follows:

\begin{itemize}
    \item We propose a gating mechanism for ranking with SID. One gate at the item level estimates each item's reliance on semantic versus collaborative signals and jointly governs fusion, alignment, and semantic granularity.

    \item Built on this gate, we design GFSA, GRCA, and GRSR to inject semantic signals into attention, modulate alignment between the semantic and collaborative spaces, and select the SID granularity for each item.

    \item We evaluate GateSID on a large-scale industrial dataset, the public Taobao-MM benchmark, and a one-month online A/B test, showing balanced industrial gains and consistent new-item advantages on the public benchmark.
    \item In addition, we conduct a systematic study of SID in ranking, covering embedding types, codebook settings, fusion strategies, and alignment schemes, and distill insightful findings. 
\end{itemize}

\section{Related Works}
\subsection{Semantic ID Encoding}

Semantic ID (SID) converts an item's multimodal content (titles, images, descriptions) into compact discrete tokens for downstream recommendation. The core technique is vector quantization: multimodal embeddings are mapped to entries of learnable codebooks, so that semantically similar items share tokens and parameters. TIGER~\cite{rajput2023recommender} introduced RQ-VAE for recommendation, where stacked codebooks perform residual quantization and each layer encodes the residual of the previous one, producing a coarse-to-fine hierarchy. OneRec~\cite{deng2025onerec} extends this idea to unify retrieval and ranking, and QARM~\cite{gai2024qarmquantitativealignment} replaces VAE training with recursive $k$-means clustering, a cheaper non-parametric alternative. Later work addresses different weaknesses of SID construction: $R^3$VAE~\cite{wan2026r3vae} stabilizes training with reference vectors to prevent codebook collapse, Hi-SAM~\cite{pan2026hisam} disentangles semantics shared across modalities from modality-specific details, and Khrylchenko et al.~\cite{khrylchenko2026variable} learn variable-length SID with Gumbel-Softmax instead of fixing the SID length in advance. Recent tokenizer advances go further: MMQ~\cite{xu2025mmq} mixes multiple quantizers and adds behavior-aware fine-tuning so the tokens adapt to downstream tasks, PaletteID~\cite{liu2026paletteid} replaces residual codebooks with prototype composition to avoid the dependency of deep codes on prefix codes, and QuaSID~\cite{hu2026quasid} learns collision-qualified SID end to end by penalizing harmful token collisions with severity-aware margins. These methods improve SID generation, and they are complementary to GateSID: GateSID treats the tokenizer as a pluggable upstream module and studies how the ranking model should use whatever SID it receives.

\subsection{Semantic ID in Ranking Models}
Simply concatenating SID embeddings as extra features already helps~\cite{singh2024better,zheng2025enhancing}, but it ignores that the value of semantic signals varies across items: collaborative signals are abundant for popular items yet nearly absent for cold-start ones. Subsequent work couples the two signals more tightly, yet the fusion remains rigid. PCR-CA~\cite{tan2025pcr} averages the two representations with uniform contrastive alignment; COINS~\cite{zhao2025coins} gates the collaborative signal alone, without interaction between modalities; SaviorRec~\cite{yao2025saviorrec} uses bidirectional target attention but treats all items uniformly; and other studies learn joint or harmonized representations~\cite{lin2025unified,zhang2025gpr,liu2025best} under the same fixed recipe for every item. In generative recommendation, DiscRec~\cite{liu2025discrec} disentangles semantic and collaborative signals with a dual-branch module and fuses the branches with a learned gate; it confirms that the two signals carry distinct distributions, but the fusion is a per-request network output without lifecycle awareness, and it is designed for sequence generation rather than ranking.

GateSID differs from these methods in the coupling mechanism between the two signals. Rather than a fixed fusion rule, a shared gate governs, on a per-item basis, three decisions that prior work treats independently: the proportion of semantic signal in the fused attention distribution (GFSA), the strength of contrastive alignment between spaces (GRCA), and the granularity at which the SID sequence is read (GRSR). Among existing methods, COINS is architecturally closest in its use of a gate. However, the gate in COINS acts solely on the collaborative pathway and therefore cannot modulate the interaction between the two modalities, whereas the gate in GateSID operates directly on the attention distributions. Moreover, the same gate simultaneously modulates the alignment loss and the SID resolution, ensuring that the three decisions remain mutually consistent rather than being optimized in isolation.
\section{Methodology}
\begin{figure*}[h]
  \centering
   
  \includegraphics[width=\linewidth]{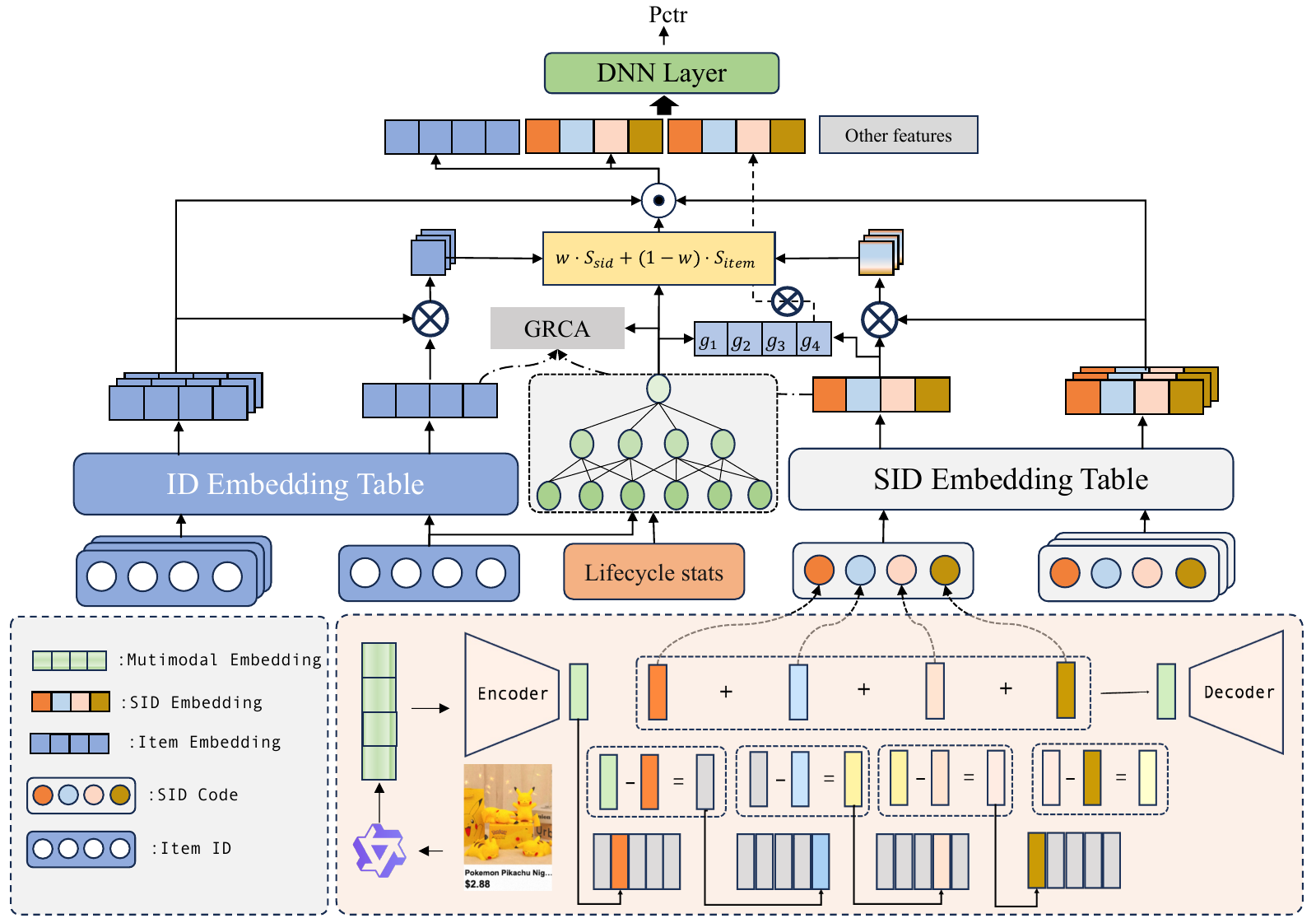}
  \caption{Overview of the GateSID framework. The bottom block constructs hierarchical SID tokens from multimodal content, while the upper ranking model consumes item and SID embeddings for prediction. A shared gate derived from item and lifecycle signals coordinates three ranking-side decisions: GFSA fuses SID and item attention distributions, GRCA weights contrastive alignment, and GRSR applies a soft prefix mask $\{g_k\}$ to select the SID granularity.}
  \label{fig:overview}
\end{figure*}

Figure~\ref{fig:overview} shows the overall architecture of GateSID. The bottom block constructs hierarchical SID offline with a residual quantized VAE. The upper block is the ranking model, where a single gate, computed from the item embedding and lifecycle statistics, controls three modules: \emph{Gating-fused shared attention} mixes the SID and item attention distributions with the gate weight; \emph{Gate-regulated contrastive alignment} scales the contrastive loss by the same gate; and \emph{Gate-regulated semantic resolution} converts the gate into a soft prefix mask over SID levels.

\subsection{Semantic ID Construction}
GateSID operates entirely in the ranking stage and consumes the SID produced by an upstream tokenizer. As a demonstration, we instantiate this upstream step with RQ-VAE~\cite{zheng2024adapting}, the widely adopted framework for SID construction. Qwen-VL~\cite{wang2024qwen2} encodes each item's textual and visual content into a multimodal embedding, and RQ-VAE then discretizes it into a hierarchical token sequence via residual quantization over stacked codebooks. Formally, the Semantic ID of item $i$ is
\begin{equation}
\label{eq:sid_def}
\mathbf{S}_i = \langle s_{i,1}, s_{i,2}, s_{i,3}, s_{i,4} \rangle, \quad s_{i,k} \in \{1, 2, \dots, 256\},
\end{equation}
where $s_{i,k}$ denotes the index in the $k$-th codebook; we use 4 residual layers with codebook size 256. Each token is mapped via an independent embedding matrix $\mathbf{E}_k$ and concatenated for the ranking model:
\begin{equation}
\label{eq:sid_concat}
\mathbf{e}_i^{\text{sid}} = \text{Concat} \left( \left[ \mathbf{E}_1(s_{i,1}) ,\mathbf{E}_2(s_{i,2}) ,\mathbf{E}_3(s_{i,3}),\mathbf{E}_4(s_{i,4}) \right] \right)
\end{equation}

\subsection{Gating-Fused Shared Attention (GFSA)}

Existing methods rely on static averaging or cross-modal interaction. Injecting semantic embeddings directly can contaminate the collaborative signals of popular items with semantic noise. GFSA instead operates at the \emph{attention-distribution level}: rather than blending embeddings, it fuses the attention weights that decide which historical items receive more focus. The fusion weight $w_i$ is computed from the collaborative embedding and statistical features (online duration, one-week exposure, and one-week clicks).
 
\begin{equation}
w_i = \sigma\left( \text{MLP}([e_i^{item} \oplus \mathcal{F}_i^{stat}]) \right)
\end{equation}
where $e_i^{item}$ is the item embedding, $\mathcal{F}_i^{stat}$ represents the statistical features, $\sigma$ is the sigmoid function, and $\oplus$ denotes the concatenation operator. These statistics (online duration, exposure, clicks) are generic lifecycle signals available on virtually any content platform rather than platform-specific fields; since the gate also conditions on the item embedding, $\mathcal{F}_i^{stat}$ can be swapped for other maturity proxies without changing the architecture, so our notion of item maturity is not tied to the industrial setting.

Let the user's SID sequence be $\mathcal{H}^{\text{sid}} = \{ \mathbf{e}_{i_1}^{\text{sid}}, \mathbf{e}_{i_2}^{\text{sid}}, \dots, \mathbf{e}_{i_L}^{\text{sid}} \} \in \mathbb{R}^{L \times d}$. GFSA first computes intra-modal attention distributions independently within $\mathcal{H}^{\text{sid}}$ and $\mathcal{H}^{\text{item}}$:
\begin{equation}
    S_{\text{sid}} = \text{Softmax}\left( \frac{(\mathbf{e}_{\text{target}}^{\text{sid}} \mathbf{W}_{Q'})(\mathcal{H}^{\text{sid}} \mathbf{W}_{K'})^\top}{\sqrt{d}} \right) 
\end{equation}

\begin{equation}
    S_{\text{item}} = \text{Softmax}\left( \frac{(\mathbf{e}_{\text{target}}^{\text{item}} \mathbf{W}_{Q'})(\mathcal{H}^{\text{item}} \mathbf{W}_{K'})^\top}{\sqrt{d}} \right) 
\end{equation}

Then, the weight $w_i$ produced by the gating network combines the two distributions into a single fused attention distribution:
\begin{equation}
\label{eq:gfsa_fuse}
    S_{\text{fused}} = w_i \cdot S_{\text{sid}} + (1 - w_i) \cdot S_{\text{item}}.
\end{equation}
Note that GFSA does not use explicit cross attention (e.g., SaviorRec), where queries from one modality attend to keys and values of another. GateSID applies the shared fused distribution $S_{\text{fused}}$ to both the $\mathcal{H}^{\text{sid}}$ and $\mathcal{H}^{\text{item}}$ sequences:
\begin{equation}
    \mathbf{h}_{\text{sid}} = S_{\text{fused}} \mathcal{H}^{\text{sid}}, \quad \mathbf{h}_{\text{item}} = S_{\text{fused}} \mathcal{H}^{\text{item}}.
\end{equation}
The two outputs are concatenated and fed into the ranking layer. Algorithm~\ref{alg:gfsa} summarizes the full GFSA pipeline.

\begin{algorithm}[t]
\caption{Gating-Fused Shared Attention (GFSA)}\label{alg:gfsa}
\begin{algorithmic}[1]
\REQUIRE $e_i^{\text{item}} \in \mathbb{R}^{d}$, $\mathcal{F}_i^{\text{stat}} \in \mathbb{R}^{d_f}$, $\mathcal{H}^{\text{sid}}, \mathcal{H}^{\text{item}} \in \mathbb{R}^{L \times d}$, $\mathbf{e}_{\text{target}}^{\text{sid}}, \mathbf{e}_{\text{target}}^{\text{item}} \in \mathbb{R}^{d}$
\STATE \textbf{Step 1: Compute gate weight}
\STATE $w_i \leftarrow \sigma(\text{MLP}([e_i^{\text{item}} \oplus \mathcal{F}_i^{\text{stat}}]))$ \hfill $\triangleright$ $(d{+}d_f) \to \mathbb{R}$
\STATE \textbf{Step 2: Compute intra-modal attention distributions}
\STATE $S_{\text{sid}} \leftarrow \text{Softmax}\left(\frac{(\mathbf{e}_{\text{target}}^{\text{sid}} \mathbf{W}_{Q'})(\mathcal{H}^{\text{sid}} \mathbf{W}_{K'})^\top}{\sqrt{d}}\right)$ \hfill $\triangleright$ $\mathbb{R}^{1 \times L}$
\STATE $S_{\text{item}} \leftarrow \text{Softmax}\left(\frac{(\mathbf{e}_{\text{target}}^{\text{item}} \mathbf{W}_{Q'})(\mathcal{H}^{\text{item}} \mathbf{W}_{K'})^\top}{\sqrt{d}}\right)$ \hfill $\triangleright$ $\mathbb{R}^{1 \times L}$
\STATE \textbf{Step 3: Fuse attention distributions}
\STATE $S_{\text{fused}} \leftarrow w_i \cdot S_{\text{sid}} + (1 - w_i) \cdot S_{\text{item}}$ \hfill $\triangleright$ $\mathbb{R}^{1 \times L}$
\STATE \textbf{Step 4: Apply fused attention to both sequences}
\STATE $\mathbf{h}_{\text{sid}} \leftarrow S_{\text{fused}} \mathcal{H}^{\text{sid}}$ \hfill $\triangleright$ $\mathbb{R}^{1 \times d}$
\STATE $\mathbf{h}_{\text{item}} \leftarrow S_{\text{fused}} \mathcal{H}^{\text{item}}$ \hfill $\triangleright$ $\mathbb{R}^{1 \times d}$
\STATE \textbf{Output}: $[\mathbf{h}_{\text{sid}} \oplus \mathbf{h}_{\text{item}}]$ for ranking layer \hfill $\triangleright$ $\mathbb{R}^{2d}$
\end{algorithmic}
\end{algorithm}

\subsection{Gate-Regulated Contrastive Alignment (GRCA)}
Standard contrastive alignment \cite{wei2021contrastive,wang2023collaborative} applies the same supervision strength to every item. We instead scale the alignment loss per item with the gate value. For a mini-batch $\mathcal{B}$, we treat the SID embedding $e_i^{sid}$ and the collaborative embedding $e_i^{item}$ of the same item $i$ as a positive pair:
\begin{equation}
\label{eq:cl_loss}
\ell_{cl}^{(i)} = -\log \frac{\exp(\text{sim}(e_i^{sid}, e_i^{item}) / \tau)}{\sum_{j \in \mathcal{B}_i} \exp(\text{sim}(e_i^{sid}, e_j^{item}) / \tau)}
\end{equation}
where $\mathcal{B}_i = \{i\} \cup \{j \in \mathcal{B} \setminus \{i\} : \mathbf{S}_j \neq \mathbf{S}_i\}$, i.e., negatives whose SID is completely identical to item $i$'s are excluded from the denominator.
The need for alignment varies across items: items with weaker collaborative evidence typically require a stronger semantic to behavioral mapping, while items with reliable collaborative structure should be less constrained by alignment. We therefore weight the contrastive loss with the gate value $w_i$:
\begin{equation}
\mathcal{L}_{cl} = \frac{1}{|\mathcal{B}|} \sum_{i \in \mathcal{B}}  w_i \cdot \ell_{cl}^{(i)}
\end{equation}
The final training objective integrates this regulated alignment loss with the primary ranking loss $\mathcal{L}_{rank}$:
\begin{equation}
\mathcal{L}_{total} = \mathcal{L}_{rank} + \lambda \mathcal{L}_{cl}
\end{equation}
where $\lambda$ is a balancing coefficient. When $w_i$ is large, the model treats the item as more semantics-reliant and aligns the two spaces more strongly; when $w_i$ is small, the constraint is relaxed and the collaborative signal is left intact. We scale the loss with the linear multiplier $w_i$ rather than nonlinear mappings such as $w_i^2$ or $\log(1+w_i)$. The linear form is the simplest choice, and it also performs best in our experiments. Since negatives are drawn in-batch, items sharing shallow SID prefixes can act as hard false negatives; this is limited because identical-SID items are masked out of the denominator and deeper SID levels separate items that share only a coarse prefix. Weighting by $w_i$ might seem to let the model shrink $\mathcal{L}_{cl}$ by driving the gate to zero, but $w_i$ is shared with GFSA and GRSR so its gradient is dominated by the ranking loss, and empirically it stays near 0.45 even for mature items (Figure~\ref{fig:sid_weight}) rather than collapsing; the weighting therefore modulates alignment strength without biasing the gate.

\subsection{Gate-Regulated Semantic Resolution (GRSR)}\label{sec:grsr}
GFSA and GRCA use the gate to decide how much semantic information to use; GRSR answers a complementary question: at what granularity each SID should be read. Recall from Eq.~\eqref{eq:sid_def} that a SID is a coarse-to-fine residual sequence $\langle s_{i,1},\dots,s_{i,K}\rangle$ with $K{=}4$. Shallow tokens are shared by many items, so their embeddings are estimated from abundant collaborative gradients; deep tokens are nearly item-specific and require reliable item-specific evidence. The useful depth therefore follows the reliability implied by the gate: items with high semantic reliance are better served by coarse, heavily shared prefixes, while items with more reliable collaborative evidence can use deeper SID levels.

GRSR realizes this adaptive depth with the shared gate $w_i$, introducing no new parameters or features. Since a larger $w_i$ means stronger reliance on semantic signals, we convert it into a behavioral maturity score:
\begin{equation}
 m_i = 1 - w_i \in [0,1].
\end{equation}
We then assign each SID level a fixed opening threshold:
\begin{equation}
\tau_k = \frac{k-1}{K}, \qquad k = 1,\dots,K.
\end{equation}
The first SID level is always used because it is the most shared and stable semantic prefix. Deeper levels are opened smoothly only when the behavioral maturity score passes their thresholds:
\begin{equation}
\label{eq:grsr_gate}
g_{i,1}=1, \qquad
g_{i,k} = \sigma\!\left( \frac{K(m_i - \tau_k)}{T} \right), \quad k=2,\dots,K,
\end{equation}
where the factor $K$ maps the normalized maturity score back to the SID-depth scale, and $T$ controls the sharpness of the transition in units of SID levels. As $T \to 0$, this becomes a hard prefix rule: level $k$ is used only after $m_i$ exceeds $\tau_k$. With a finite $T$, the transition remains smooth and differentiable. Because $\tau_1 < \tau_2 < \dots < \tau_K$, deeper SID levels require a higher behavioral maturity score to activate, producing a monotonic soft prefix mask.  The resolution gated SID embedding replaces the plain concatenation of Eq.~\eqref{eq:sid_concat}:
\begin{equation}
\label{eq:sid_gated}
\mathbf{e}_i^{\text{sid}} = \text{Concat}\!\left[\, g_{i,1}\,\mathbf{E}_1(s_{i,1}),\; g_{i,2}\,\mathbf{E}_2(s_{i,2}),\; \dots,\; g_{i,K}\,\mathbf{E}_K(s_{i,K}) \,\right].
\end{equation}
Equation~\eqref{eq:sid_concat} is the special case where every $g_{i,k}{=}1$, so GRSR generalizes rather than contradicts the plain concatenation, and the ``w/o GRSR'' ablation is exactly this fallback. The level-wise mask does assume a residual hierarchy: for parallel codebooks (e.g., OPQ/PQ) or prototype composition without a prefix order, GFSA and GRCA still apply while GRSR would need an alternative level ordering.

\setlength{\textfloatsep}{6pt plus 1pt minus 2pt}
\setlength{\floatsep}{6pt plus 1pt minus 2pt}
\setlength{\intextsep}{6pt plus 1pt minus 2pt}
\begin{table*}[!t]
\centering
\small
  \caption{AUC(\%,$\uparrow$) and GAUC(\%,$\uparrow$) results of GateSID and baselines on different item groups. Cold-start items are those that have been online for fewer than 20 days. Popular items are those that have been online for more than 300 days. All GateSID improvements are statistically significant under a two-sided $t$-test, with $p$-values ranging from $p < 0.01$ for popular-item gains to $p < 0.05$ for cold-start margins.}
\label{tab:full_comparison}
\resizebox{\textwidth}{!}{%
\renewcommand{\arraystretch}{1}%
\begin{tabular}{@{}lllcccccccccc@{}}
\toprule
\textbf{Group} & \textbf{Task} & \textbf{Metric} & \textbf{Base} & \textbf{COINS} & \textbf{PCR-CA} & \textbf{URL4DR} & \textbf{SPM-SID} & \textbf{SaviorRec} & \textbf{QARM} & \textbf{GateSID} & \textbf{Improv.} \\
\midrule
\multirow{4}{*}{Overall}
& \multirow{2}{*}{CTR}  & AUC   & 0.6953 & 0.6980 & 0.6977 & 0.6977 & 0.6955 & \underline{0.6984} & 0.6981 & \textbf{0.7011}$^{*}$ & +0.27\% \\
&                       & GAUC  & 0.6562 & 0.6572 & 0.6570 & 0.6591 & 0.6567 & \underline{0.6591} & 0.6588 & \textbf{0.6617}$^{*}$ & +0.26\% \\
\cmidrule(lr){2-12}
& \multirow{2}{*}{CTCVR} & AUC   & 0.8449 & 0.8477 & 0.8472 & 0.8469 & 0.8459 & 0.8480 & \underline{0.8481} & \textbf{0.8506}$^{*}$ & +0.25\% \\
&                         & GAUC  & 0.7129 & 0.7136 & 0.7145 & 0.7159 & 0.7144 & 0.7192 & \underline{0.7195} & \textbf{0.7231}$^{*}$ & +0.36\% \\
\midrule
\multirow{4}{*}{Popular}
& \multirow{2}{*}{CTR}  & AUC   & 0.6965 & 0.6969 & 0.6968 & 0.6970 & 0.6966 & 0.6971 & \underline{0.6972} & \textbf{0.7007}$^{*}$ & +0.42\% \\
&                       & GAUC  & 0.6221 & 0.6225 & 0.6224 & 0.6228 & 0.6222 & \underline{0.6230} & 0.6229 & \textbf{0.6264}$^{*}$ & +0.34\% \\
\cmidrule(lr){2-12}
& \multirow{2}{*}{CTCVR} & AUC   & 0.8463 & 0.8469 & 0.8467 & 0.8470 & 0.8465 & \underline{0.8472} & 0.8471 & \textbf{0.8509}$^{*}$ & +0.37\% \\
&                         & GAUC  & 0.6483 & 0.6487 & 0.6485 & 0.6489 & 0.6484 & \underline{0.6492} & 0.6491 & \textbf{0.6544}$^{*}$ & +0.52\% \\
\midrule
\multirow{4}{*}{Cold-start}
& \multirow{2}{*}{CTR}  & AUC   & 0.6829 & 0.6853 & 0.6847 & 0.6852 & 0.6832 & 0.6853 & \underline{0.6864} & \textbf{0.6884}$^{*}$ & +0.20\% \\
&                       & GAUC  & 0.5967 & 0.5971 & \underline{0.5984} & 0.5985 & 0.5959 & 0.5978 & 0.5981 & \textbf{0.6003}$^{*}$ & +0.19\% \\
\cmidrule(lr){2-12}
& \multirow{2}{*}{CTCVR} & AUC   & 0.8301 & 0.8336 & 0.8328 & 0.8327 & 0.8316 & 0.8337 & \underline{0.8346} & \textbf{0.8365}$^{*}$ & +0.19\% \\
&                         & GAUC  & 0.6218 & 0.6250 & 0.6266 & 0.6253 & 0.6254 & \underline{0.6297} & 0.6286 & \textbf{0.6321}$^{*}$ & +0.24\% \\
\bottomrule
\end{tabular}%
}
\end{table*}

\section{Experiments}
\subsection{Experiments Settings}
\subsubsection{Datasets}
We evaluate on two datasets. The main evaluation uses a large-scale industrial dataset with over 1 billion production logs; following the online setup, we split the data by time and reserve the most recent day for validation and testing, and we consider two tasks, click (CTR) and pay (CTCVR), reporting AUC and GAUC for both. To examine generality beyond our production traffic, we also evaluate on the public Taobao-MM dataset~\cite{wu2025muse}, which provides long user behavior sequences and multimodal item embeddings. We follow the PaletteID~\cite{liu2026paletteid} protocol and evaluate with both DIN and RankMixer~\cite{zhu2025rankmixer} as backbones for click prediction.

\subsubsection{Baselines}
We compare GateSID with recent SID-enhanced baselines: COINS~\cite{zhao2025coins}, PCR-CA~\cite{tan2025pcr}, URL4DR~\cite{lin2025unified}, SPM-SID~\cite{singh2024better}, SaviorRec~\cite{yao2025saviorrec}, and QARM~\cite{gai2024qarmquantitativealignment}. All models share the same data partitions and training protocol.
\subsubsection{Implementation Details}
The RQ-VAE uses 4 residual layers, each with a codebook of size 256 and embedding dimension 64, trained for 10 epochs with batch size 4096. The ranking model employs AdamW with batch size 4096, SID embeddings of dimension 32, a balancing coefficient $\lambda=0.1$, and a 3-layer ReLU MLP for joint CTR and CTCVR prediction throughout training. The GRCA contrastive loss uses in-batch negatives sampled across users within each batch, and negatives whose SID is completely identical to the anchor are excluded. For GRSR, the soft-prefix temperature is $T=1.0$. The statistical features $\mathcal{F}_i^{stat}$ (online duration, one-week exposure, one-week clicks) are updated daily in the production system. The gate is robust to this staleness because it also conditions on item embeddings, which are always current. 
\subsection{Overall Performance}
Table~\ref{tab:full_comparison} compares GateSID with all baselines. Every method that uses SID improves over the base model, which confirms that semantic information helps ranking. GateSID leads the strongest baseline on every metric, by +0.27\% CTR AUC and +0.36\% CTCVR GAUC on the overall traffic. For a production ranking model refined over years, margins of this magnitude are meaningful.

The breakdown by item group shows where these gains come from. On popular items, GateSID leads the strongest baseline by +0.52\% CTCVR GAUC: the other methods inject semantic embeddings uniformly, diluting the reliable collaborative signals of popular items, whereas the gate reduces the semantic weight exactly where collaborative evidence is strong. On cold-start items, GateSID improves over the base model by +0.55\% CTR AUC and +0.64\% CTCVR AUC, so semantics does compensate for missing interactions; its margin over the baselines is smaller here (+0.19\% CTCVR AUC), as expected, since when collaborative evidence is absent every SID method already leans on semantics. GateSID thus improves both ends of the item lifecycle at once, rather than trading one off against the other, with its widest advantage on popular items, where a fixed fusion rule would otherwise dilute the reliable collaborative signals.

\subsection{Results on the Public Taobao-MM Dataset}
To check that our findings are not specific to the production system, we also evaluate on Taobao-MM with two backbones, DIN and RankMixer. Table~\ref{tab:taobao_mm} reports both overall results and item group breakdowns. With DIN, GateSID achieves the best overall performance, leading the strongest baseline COINS by +0.68\% AUC and +0.64\% GAUC. The gain is larger on new items: GateSID improves over the strongest baseline by +0.99\% AUC and +0.99\% GAUC. On old items, COINS is slightly better than GateSID by 0.21\% AUC and 0.21\% GAUC, but both methods clearly outperform the base model, suggesting that semantic content remains useful even for  old items.

The RankMixer results show the same pattern with a stronger overall advantage: GateSID leads COINS by +1.00\% AUC and +0.96\% GAUC overall, leads the best non-GateSID method on new items by +0.37\% AUC and +0.32\% GAUC, and on old items by +0.18\% AUC and +0.19\% GAUC. This clarifies that the old group in Taobao-MM is not equivalent to the highly reliable popular items in production, since semantics still brings large gains there. More importantly, GateSID gives the best new-item performance under both backbones while remaining competitive or best on old items, supporting our claim that adaptive SID usage balances the two signals across item groups.

\begin{table}[!tbp]
\centering
\small
\caption{Results on Taobao-MM with the DIN and RankMixer backbones (CTR task). New items have at most 50 training exposures and old items have more than 100. We report AUC and GAUC on the overall test set and on new/old item groups.}
\label{tab:taobao_mm}
\resizebox{\columnwidth}{!}{%
\begin{tabular}{@{}lllccccc@{}}
\toprule
\textbf{Backbone} & \textbf{Group} & \textbf{Metric} & \textbf{Base} & \textbf{SaviorRec} & \textbf{COINS} & \textbf{PCR-CA} & \textbf{GateSID} \\
\midrule
\multirow{6}{*}{DIN}
& \multirow{2}{*}{Overall} & AUC  & 0.5912 & 0.5928 & \underline{0.6062} & 0.5927 & \textbf{0.6130} \\
&                          & GAUC & 0.5898 & 0.5911 & \underline{0.6050} & 0.5909 & \textbf{0.6114} \\
\cmidrule(lr){2-8}
& \multirow{2}{*}{New}     & AUC  & 0.5604 & \underline{0.5642} & 0.5627 & 0.5627 & \textbf{0.5741} \\
&                          & GAUC & 0.5598 & \underline{0.5631} & 0.5622 & 0.5617 & \textbf{0.5730} \\
\cmidrule(lr){2-8}
& \multirow{2}{*}{Old}     & AUC  & 0.6126 & 0.6122 & \textbf{0.6395} & 0.6131 & \underline{0.6374} \\
&                          & GAUC & 0.6113 & 0.6108 & \textbf{0.6382} & 0.6116 & \underline{0.6361} \\
\midrule
\multirow{6}{*}{RankMixer}
& \multirow{2}{*}{Overall} & AUC  & 0.5901 & 0.5914 & \underline{0.6008} & 0.5912 & \textbf{0.6108} \\
&                          & GAUC & 0.5887 & 0.5901 & \underline{0.5996} & 0.5898 & \textbf{0.6092} \\
\cmidrule(lr){2-8}
& \multirow{2}{*}{New}     & AUC  & 0.5621 & \underline{0.5630} & 0.5628 & 0.5626 & \textbf{0.5667} \\
&                          & GAUC & 0.5613 & \underline{0.5622} & 0.5620 & 0.5619 & \textbf{0.5654} \\
\cmidrule(lr){2-8}
& \multirow{2}{*}{Old}     & AUC  & 0.6095 & 0.6106 & \underline{0.6351} & 0.6103 & \textbf{0.6369} \\
&                          & GAUC & 0.6085 & 0.6097 & \underline{0.6338} & 0.6094 & \textbf{0.6357} \\
\bottomrule
\end{tabular}%
}
\end{table}

\subsection{Fusion Operator Analysis}
\begin{figure}[!tbp]
\centering
\includegraphics[width=\columnwidth]{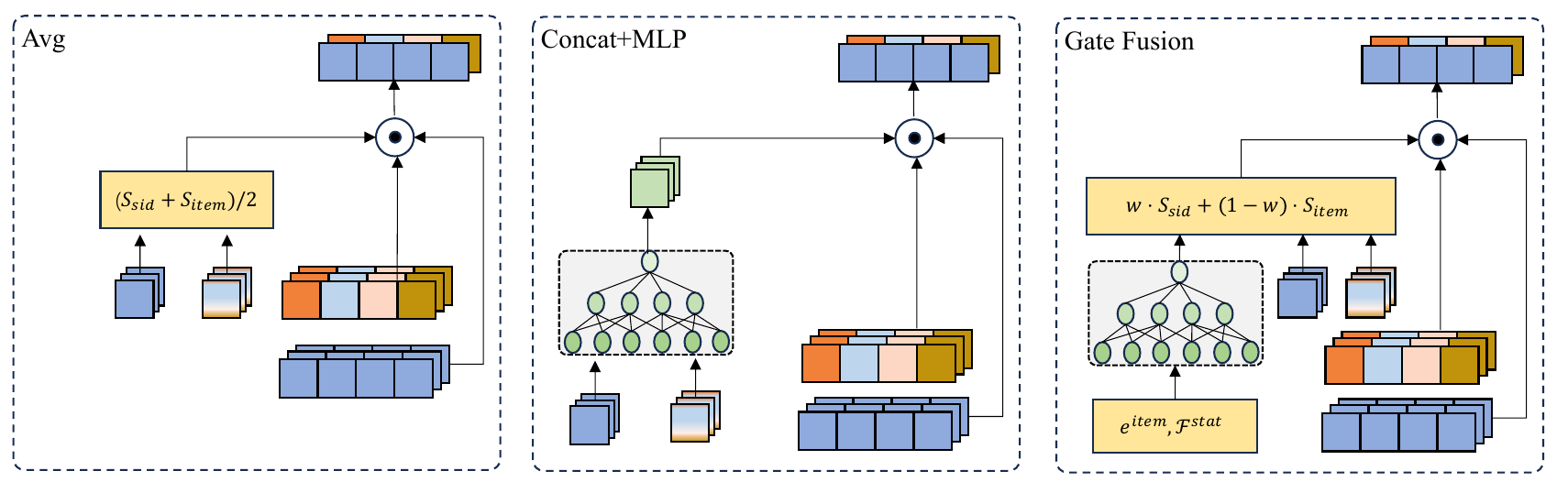}
\caption{Illustration of different fusion operators: Average, Concat + MLP, and Gate Fusion.}
\label{fig:fusion_operator}
\end{figure}

We compare four fusion operators under the same GFSA attention, varying only how the semantic and collaborative signals are aggregated. In every variant the gate $w_i$ is still computed and still drives GRCA and GRSR; only the aggregation rule in Eq.~\eqref{eq:gfsa_fuse} is replaced, so the comparison isolates the fusion operator itself rather than the presence of the gate. \textbf{Average} assigns equal weight to the two signals. \textbf{Static Weighted} applies one global weight to all items, tuned on the validation set over $\{0.3, 0.4, 0.5, 0.6\}$; the grid contains $0.5$, which reduces to Average, and the selected value $0.4$ sits close to the population mean of the learned gates. \textbf{Concat + MLP} concatenates the two signals and projects them through a feedforward network, which adds capacity but no explicit weighting. \textbf{Gate Fusion} is our adaptive operator, which produces one weight per item. Figure~\ref{fig:fusion_operator} illustrates these designs.

Table~\ref{tab:fusion_operator} reports the results. The three static operators fall within 0.10\% CTR AUC of each other: tuning one global weight (Static Weighted) barely improves on plain averaging, and Concat + MLP adds capacity without changing the picture. This narrow spread is itself informative: no single weight, and no amount of capacity applied uniformly, can express the balance that individual items need. Gate Fusion breaks out of this band and ranks first on every metric, ahead of the strongest static variant by +0.27\% CTR AUC and +0.45\% CTCVR GAUC. Since Static Weighted is the best fixed weighting, this margin isolates the value of adapting the weight per item rather than the value of weighting itself, and it is widest on CTCVR GAUC, the metric most exposed to behavioral sparsity.

The learned gate (Figure~\ref{fig:sid_weight}) behaves as expected: it assigns larger weights to items with less reliable collaborative evidence and shifts toward the collaborative path when behavior becomes more informative. Notably, the gate does not saturate to zero even for mature items, settling around 0.45. We view this as the optimum the gate discovers rather than a failure to saturate. All methods that use SID outperform the purely collaborative base model even on popular items (Table~\ref{tab:full_comparison}), so semantic content remains useful throughout the lifecycle; and the best fixed weighting (Static Weighted, $w{=}0.4$) still trails the adaptive gate by +0.27\% CTR AUC, so no constant weight can reproduce this balance.

We further examine the effect of gate input in the lower panel of Table~\ref{tab:fusion_operator}. Using either $e^{item}$ or $\mathcal{F}^{stat}$ alone underperforms their concatenation, confirming that both item identity and statistical features carry complementary information for gating. Concretely, the concatenated input reaches 0.7011 CTR AUC and 0.7231 CTCVR GAUC, ahead of the stronger single input (only $e^{item}$) by +0.25\% on both. The two inputs fail in complementary ways: identity alone cannot see how much behavioral evidence an item has accumulated, while statistics alone cannot distinguish items with similar exposure. Only together do they let the gate separate a well-observed popular item from a rarely-observed one.

\begin{table}[!tbp]
\centering
\small
\caption{Comparison of fusion operators and ablation of gate inputs with the attention mechanism fixed to GFSA.}
\label{tab:fusion_operator}
\begin{tabular}{@{}llcccc@{}}
\toprule
& & \multicolumn{2}{c}{CTR} & \multicolumn{2}{c}{CTCVR} \\
\cmidrule(lr){3-4} \cmidrule(lr){5-6}
\textbf{Category} & \textbf{Variant} & AUC & GAUC & AUC & GAUC \\
\midrule
\multirow{4}{*}{Fusion}
& Average & 0.6974 & 0.6581 & 0.8472 & 0.7172 \\
& Static Weighted & 0.6984 & 0.6590 & 0.8480 & 0.7186 \\
& Concat + MLP & 0.6978 & 0.6583 & 0.8478 & 0.7173 \\
& \textbf{Gate Fusion} & \textbf{0.7011} & \textbf{0.6617} & \textbf{0.8506} & \textbf{0.7231} \\
\midrule
\multirow{2}{*}{Gate Input}
& only $e^{item}$ & 0.6986 & 0.6598 & 0.8489 & 0.7206 \\
& only $\mathcal{F}^{stat}$ & 0.6981 & 0.6590 & 0.8484 & 0.7184 \\
\bottomrule
\end{tabular}
\end{table}

\subsection{Embedding Type Analysis}
Having validated the efficacy of Gate Fusion in integrating semantic and collaborative signals, we next investigate the impact of different semantic embedding formulations. Table~\ref{tab:sid_structural_gain} presents an ablation study that isolates the embedding types and attention.

Switching from the original attention to GFSA with the same MM embedding improves CTR AUC by +0.20\% and CTCVR AUC by +0.10\%, showing that cross-modal attention fusion is effective independent of the embedding type. Under the same GFSA, SID outperforms MM across all four metrics; the gap is largest in CTCVR GAUC at +0.46\%, suggesting that discrete tokenization is especially helpful for conversion prediction where behavioral sparsity is severe. Combining MM and SID yields the best results, outperforming SID alone by +0.16\% CTR AUC and MM alone by +0.36\% CTR AUC.

\begin{figure}[!tbp]
    \centering
    \begin{subfigure}[b]{0.48\linewidth}
        \centering
        \includegraphics[width=\linewidth]{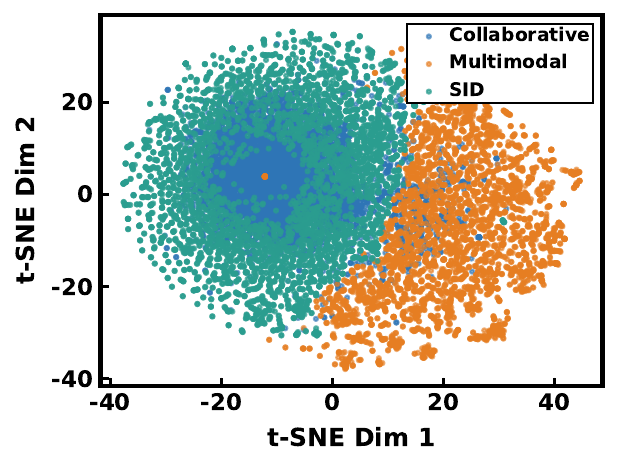}
        \caption{Distributions of three types of embedding.}
        \label{fig:sid_case}
    \end{subfigure}
    \hfill
    \begin{subfigure}[b]{0.48\linewidth}
        \centering
        \includegraphics[width=\linewidth]{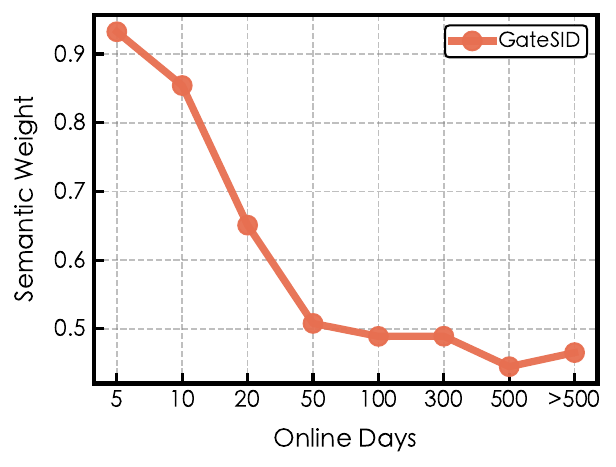}
        \caption{Gate weight for items with different online days.}
        \label{fig:sid_weight}
    \end{subfigure}
    \caption{Comparison and visualization of different embedding representations.}
    \label{fig:sid_embedding_comparison}
\end{figure}

Shared codebook assignments create implicit category groupings that regularize the representation space, and fine-tuning lets SID token embeddings absorb collaborative signals during ranking. Both effects should move SID embeddings toward the collaborative space, and a t-SNE projection of 10,000 items confirms this (Figure~\ref{fig:sid_case}): the centroid distance between SID and item embeddings is 1.24, versus 1.89 between MM and item embeddings, and the MMD between the two distributions drops from 0.186 to 0.073. SID also has a practical advantage in dimensionality: fixed multimodal embeddings must be compressed to 32 dimensions for online serving, which loses information, whereas SID represents each item as a short token sequence whose embeddings are fine-tuned jointly, preserving richer structure in compact form.

\begin{table}[!tbp]
\centering
\small
\caption{The AUC(\%,$\uparrow$) and GAUC(\%,$\uparrow$) results of ablation study of embedding types and attention mechanisms.}
\label{tab:sid_structural_gain}
\begin{tabular}{@{}llcccc@{}}
\toprule
\textbf{Embedding} & \textbf{Attention} & \multicolumn{2}{c}{CTR} & \multicolumn{2}{c}{CTCVR} \\
\cmidrule(lr){3-4} \cmidrule(l){5-6}
& & AUC & GAUC & AUC & GAUC \\
\midrule
Item & Origin & 0.6953 & 0.6562 & 0.8449 & 0.7129 \\
Item + MM & Origin & 0.6971 & 0.6581 & 0.8470 & 0.7152 \\
Item + MM & GFSA & 0.6991 & 0.6591 & 0.8480 & 0.7185 \\
Item + SID & GFSA & 0.7011 & 0.6617 & 0.8506 & 0.7231 \\
\textbf{Item + MM + SID} & \textbf{GFSA} & \textbf{0.7027} & \textbf{0.6642} & \textbf{0.8529} & \textbf{0.7257} \\
\bottomrule
\end{tabular}
\end{table}

\subsection{Semantic Code Analysis}
We next study how SID hyperparameters affect ranking. The two main ones are the codebook size and the number of layers, which jointly control the capacity of the SID space.

\begin{figure}[!tbp]
\centering
\includegraphics[width=\columnwidth]{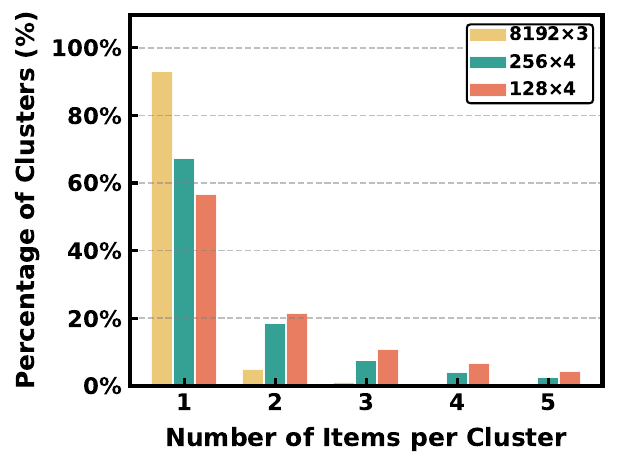}
\caption{Cluster size distributions under different SID configurations.}
\label{fig:cluster_size_dist}
\end{figure}

Table~\ref{tab:sid_hyperparam} shows three findings. First, both text-only and multimodal SID representations outperform the base model; the best configuration, multimodal $128 \times 4$, improves CTR AUC by +0.49\% and CTCVR AUC by +0.60\% over the base. Second, at the same $256 \times 4$ setting, multimodal SID outperforms text-only SID by +0.09\% CTR AUC and +0.17\% CTCVR AUC, so visual content adds useful cues beyond text. Third, the two compact codebooks ($128 \times 4$ and $256 \times 4$) perform nearly the same (a 0.06\% CTR AUC gap), and both clearly outperform $8192 \times 3$, which improves over the base model by only +0.01\%.  A smaller codebook forces more items to share a code, which aggregates collaborative signals across semantically similar items and transfers knowledge from popular items to cold-start ones. Figure~\ref{fig:cluster_size_dist} supports this: shrinking the codebook concentrates more items per cluster and raises the collision rate. Large vocabularies do the opposite: they split items into fine partitions that isolate cold-start items and weaken generalization.  We note that this conclusion is specific to ranking models.

\begin{table}[!tbp]
\centering
\small
\caption{Impact of different modal representations and SID configurations on ranking performance.}
\label{tab:sid_hyperparam}
\begin{tabular}{@{}llcccc@{}}
\toprule
\textbf{Modality} & \textbf{SID Settings} & \multicolumn{2}{c}{CTR} & \multicolumn{2}{c}{CTCVR} \\
\cmidrule(lr){3-4} \cmidrule(l){5-6}
& & AUC & GAUC & AUC & GAUC \\
\midrule
Base & - & 0.6953 & 0.6562 & 0.8449 & 0.7129 \\
Text & 256$\times$4 & 0.6987 & 0.6592 & 0.8482 & 0.7199 \\
Multimodal & 8192$\times$3 & 0.6954 & 0.6566 & 0.8453 & 0.7134 \\
Multimodal & 256$\times$4 & 0.6996 & 0.6603 & 0.8499 & 0.7206 \\
Multimodal & 128$\times$4 & \textbf{0.7002} & \textbf{0.6606} & \textbf{0.8509} & \textbf{0.7210} \\
\bottomrule
\end{tabular}
\end{table}

\subsection{Ablation Study}
\begin{figure}[!tbp]
\centering
\includegraphics[width=\columnwidth]{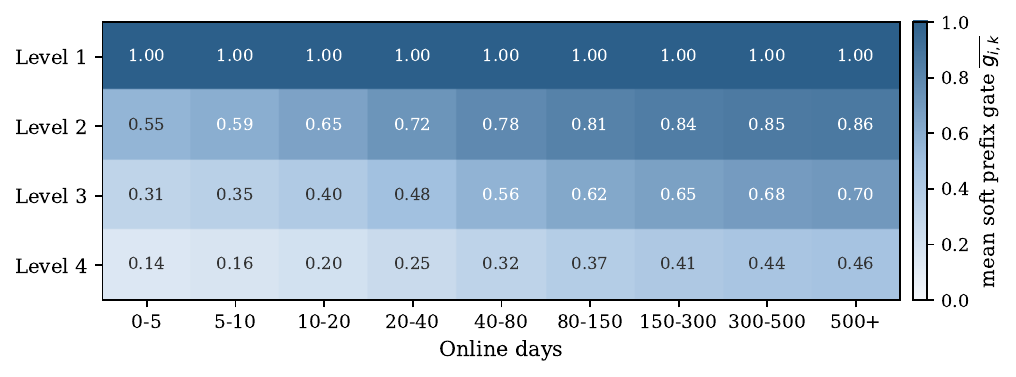}
\caption{Mean soft prefix gate $\overline{g_{i,k}}$ of each SID level across online-day buckets, computed with Eq.~\eqref{eq:grsr_gate} at $T{=}1.0$; it is a soft mask, not a normalized distribution.}
\label{fig:grsr_heatmap}
\end{figure}

Finally, we evaluate the contribution of each gate-driven component by removing one at a time from the full model. Removing a module bypasses only its own mechanism while the gate is still trained and $w_i$ keeps driving the other two: ``w/o GFSA'' disables the fusion in Eq.~\eqref{eq:gfsa_fuse} but retains $w_i$ in GRCA and GRSR, ``w/o GRCA'' drops $\mathcal{L}_{cl}$, and ``w/o GRSR'' sets every $g_{i,k}{=}1$, falling back to Eq.~\eqref{eq:sid_concat}. Each drop is thus attributable to the removed mechanism, not to losing the gate. As shown in Table~\ref{tab:alignment_comparison}, all three components matter. Removing GFSA hurts CTR AUC the most ($-0.22\%$), confirming that attention-distribution fusion is the primary carrier of semantic signals in ranking. Removing GRSR costs $-0.21\%$ CTR AUC and $-0.22\%$ CTCVR GAUC, showing that adaptively choosing the SID granularity is nearly as important as the fusion itself. Removing GRCA degrades mainly the GAUC metrics ($-0.09\%$ CTR GAUC and $-0.11\%$ CTCVR GAUC), since without gate-regulated weighting the alignment supervision cannot adapt to item maturity; notably, replacing GRCA with a uniform contrastive loss (the same loss without gate weighting) yields a CTR AUC of 0.6988, still well below the full model, which shows that the gate-regulated weighting, rather than the mere presence of contrastive supervision, drives the gain. Figure~\ref{fig:grsr_heatmap} further visualizes the learned soft prefix gates of GRSR. Level 1 is always active, while deeper levels open more gradually as the learned behavioral maturity increases. This figure isolates the resolution decision; the overall reliance on SID is still controlled by the shared gate $w_i$.

\begin{figure}[!tbp]
\centering
\includegraphics[width=\columnwidth]{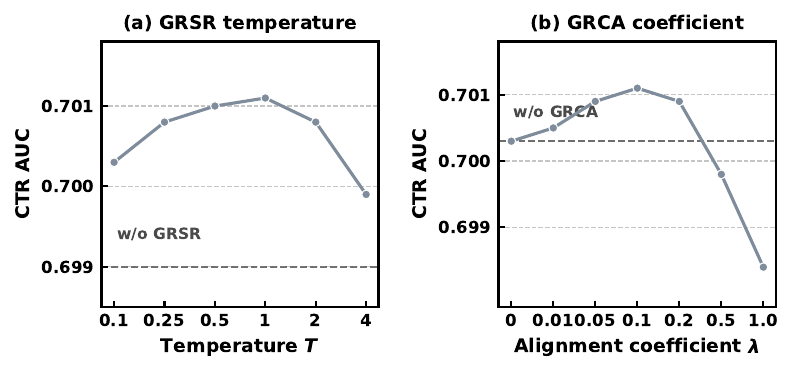}
\caption{Hyperparameter sensitivity on the industrial dataset for CTR AUC.}
\label{fig:sensitivity}
\end{figure}

We further probe how sensitive GateSID is to the key hyperparameters of GRSR and GRCA in Figure~\ref{fig:sensitivity}. For GRSR, varying the temperature $T$ from 0.1 to 4.0 moves CTR AUC by at most 0.12\%, and the curve stays within 0.03\% of the peak for $T \in [0.25, 2.0]$. Performance drops only at the extremes. As $T$ approaches zero the prefix gates approach hard truncation, and as $T$ grows every deeper-level gate converges to 0.5 regardless of the item, so GRSR loses its item-adaptive resolution and behaves like a fixed rescaling of the SID sequence. For GRCA, the coefficient $\lambda$ traces a gentle inverted-U curve that peaks at the default $\lambda = 0.1$ and stays within 0.02\% for $\lambda \in [0.05, 0.2]$. A large $\lambda$ over-aligns the two spaces: at $\lambda = 1.0$ the model falls 0.27\% below the default and even below the variant without alignment, because uniform strength supervision over-constrains items whose collaborative embeddings are already well trained. Both modules are therefore robust around their defaults, and only extreme settings break the intended behavior.

\begin{table}[!tbp]
\centering
\small
\caption{Ablation study of GateSID components. Each row removes one gate-driven component from the full model.}
\label{tab:alignment_comparison}
\begin{tabular}{@{}lcccc@{}}
\toprule
& \multicolumn{2}{c}{CTR} & \multicolumn{2}{c}{CTCVR} \\
\cmidrule(lr){2-3} \cmidrule(l){4-5}
\textbf{Variant} & AUC & GAUC & AUC & GAUC \\
\midrule
\textbf{GateSID (full)} & \textbf{0.7011} & \textbf{0.6617} & \textbf{0.8506} & \textbf{0.7231} \\
w/o GRSR & 0.6990 & 0.6597 & 0.8487 & 0.7209 \\
w/o GRCA & 0.7003 & 0.6608 & 0.8496 & 0.7220 \\
w/o GFSA & 0.6989 & 0.6599 & 0.8484 & 0.7204 \\
\bottomrule
\end{tabular}
\end{table}

\subsection{Online A/B Test}
We deployed GateSID on our production platform and ran a one-month A/B test. Following user-level randomization, 20\% of the traffic  went to the treatment group and an equal share served as control. GateSID improves GMV by +2.6\%, CTR by +1.1\%, and Order by +1.6\%, with less than 5 ms of added serving latency. Guardrail metrics (page load time, ad revenue, and diversity) remain stable.

\begin{figure}[!tbp]
    \centering
    \begin{subfigure}[b]{0.48\linewidth}
        \centering
        \includegraphics[width=\linewidth]{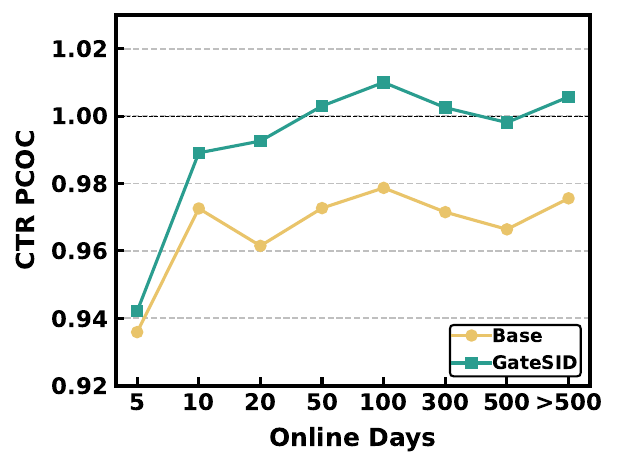}
        \caption{CTR PCOC by online days.}
        \label{fig:pcoc_by_days}
    \end{subfigure}
    \hfill
    \begin{subfigure}[b]{0.48\linewidth}
        \centering
        \includegraphics[width=\linewidth]{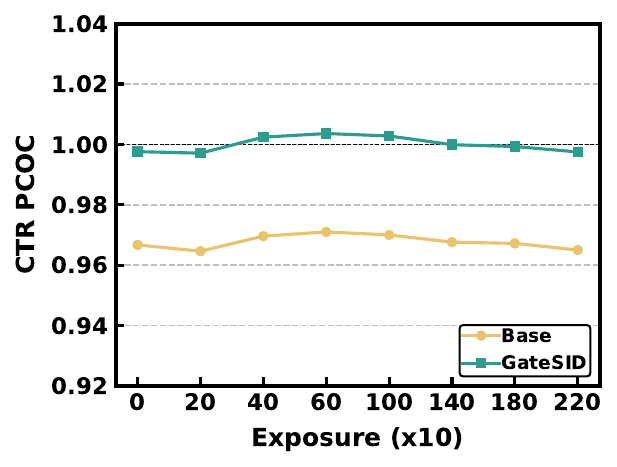}
        \caption{CTR PCOC by exposures.}
        \label{fig:pcoc_by_exposure}
    \end{subfigure}
    \caption{CTR PCOC comparison between GateSID and the baseline.}
    \label{fig:pcoc_comparison}
\end{figure}

We further examine online prediction calibration with PCOC, broken down by item age and exposure count. PCOC is the ratio of predicted to observed CTR; 1.0 means perfect calibration. Grouped by online days (Figure~\ref{fig:pcoc_by_days}), the baseline underestimates CTR in every group (PCOC below 1.0), while GateSID stays close to 1.0; the same holds when grouped by exposure count (Figure~\ref{fig:pcoc_by_exposure}). GateSID uses no explicit calibration layer, so this improvement comes from better representations rather than post-hoc correction.

\section{Conclusion}
We presented GateSID, a framework that balances collaborative and semantic signals across the item lifecycle with one shared gate per item. The same gate drives three modules: GFSA fuses the SID and item attention distributions, GRCA scales the alignment loss, and GRSR selects the SID granularity, so cold-start items receive stronger semantic support while popular items keep their collaborative signals intact. On the industrial dataset, GateSID leads the strongest baseline on every metric, with its widest margin on popular items, and a one-month online A/B test (GMV +2.6\%, CTR +1.1\%) confirms its practical value; on the public Taobao-MM benchmark it leads overall under both DIN and RankMixer backbones and gives the strongest new-item performance. Our study of SID in ranking further distills several findings: adaptive gating beats static aggregation, since no constant weight reproduces the balance that varies with item maturity; quantized SID outperforms raw multimodal embeddings, because codebook sharing regularizes the representation space and fine-tuning moves the tokens toward the collaborative space; and compact codebooks outperform large ones by concentrating items per cluster and transferring knowledge from popular to cold-start items. We hope these findings are useful for future work on ranking with semantic representations.

\section{Ethical Considerations}
Our work studies how to integrate semantic item representations into industrial ranking models. The training data consists of anon\-y\-mized interaction logs and public item content, and no personally identifiable information is used. By improving the visibility of cold-start and long-tail items, GateSID may in fact benefit content diversity and fairness toward new sellers. We do not identify any negative societal impact, potential misuse, or fairness, privacy, security, or safety concerns arising from this work.
 
\bibliographystyle{ACM-Reference-Format}
\balance
\bibliography{base}

@article{zhao2025coins,
  title={COINS: SemantiC Ids Enhanced COLd Item RepresentatioN for Click-through Rate Prediction in E-commerce Search},
  author={Zhao, Qihang and Sun, Zhongbo and Zheng, Xiaoyang and Guo, Xian and Wang, Siyuan and Liang, Zihan and Peng, Mingcan and Chen, Ben and Lei, Chenyi},
  journal={arXiv preprint arXiv:2510.12604},
  year={2025}
}

@article{wang2026sort,
  title={SORT: A Systematically Optimized Ranking Transformer for Industrial-scale Recommenders},
  author={Wang, Chunqi and Wu, Bingchao and Pang, Taotian and Wang, Jiahao and Yang, Jie and Liu, Jia and Zhang, Hao and Zhu, Hai and Shen, Lei and Wang, Shizhun and others},
  journal={arXiv preprint arXiv:2603.03988},
  year={2026}
}

@inproceedings{singh2024better,
  title={Better generalization with semantic ids: A case study in ranking for recommendations},
  author={Singh, Anima and Vu, Trung and Mehta, Nikhil and Keshavan, Raghunandan and Sathiamoorthy, Maheswaran and Zheng, Yilin and Hong, Lichan and Heldt, Lukasz and Wei, Li and Tandon, Devansh and others},
  booktitle={Proceedings of the 18th ACM Conference on Recommender Systems},
  pages={1039--1044},
  year={2024}
}

@inproceedings{zheng2025enhancing,
  title={Enhancing embedding representation stability in recommendation systems with semantic id},
  author={Zheng, Carolina and Huang, Minhui and Pedchenko, Dmitrii and Rangadurai, Kaushik and Wang, Siyu and Xia, Fan and Nahum, Gaby and Lei, Jie and Yang, Yang and Liu, Tao and others},
  booktitle={Proceedings of the Nineteenth ACM Conference on Recommender Systems},
  pages={954--957},
  year={2025}
}

@article{lin2025unified,
  title={Unified semantic and ID representation learning for deep recommenders},
  author={Lin, Guanyu and Hua, Zhigang and Feng, Tao and Yang, Shuang and Long, Bo and You, Jiaxuan},
  journal={arXiv preprint arXiv:2502.16474},
  year={2025}
}

@inproceedings{hu2025alphafuse,
  title={Alphafuse: Learn id embeddings for sequential recommendation in null space of language embeddings},
  author={Hu, Guoqing and Zhang, An and Liu, Shuo and Cai, Zhibo and Yang, Xun and Wang, Xiang},
  booktitle={Proceedings of the 48th International ACM SIGIR Conference on Research and Development in Information Retrieval},
  pages={1614--1623},
  year={2025}
}

@inproceedings{zheng2024adapting,
  title={Adapting large language models by integrating collaborative semantics for recommendation},
  author={Zheng, Bowen and Hou, Yupeng and Lu, Hongyu and Chen, Yu and Zhao, Wayne Xin and Chen, Ming and Wen, Ji-Rong},
  booktitle={2024 IEEE 40th International Conference on Data Engineering (ICDE)},
  pages={1435--1448},
  year={2024},
  organization={IEEE}
}

@inproceedings{ganhor2024multimodal,
  title={A multimodal single-branch embedding network for recommendation in cold-start and missing modality scenarios},
  author={Ganh{\"o}r, Christian and Moscati, Marta and Hausberger, Anna and Nawaz, Shah and Schedl, Markus},
  booktitle={Proceedings of the 18th ACM Conference on Recommender Systems},
  pages={380--390},
  year={2024}
}

@article{wang2023collaborative,
  title={Collaborative semantic alignment in recommendation systems},
  author={Wang, Chen and Yang, Liangwei and Liu, Zhiwei and Liu, Xiaolong and Yang, Mingdai and Liang, Yueqing and Yu, Philip S},
  journal={arXiv preprint arXiv:2310.09400},
  year={2023}
}

@article{wang2024qwen2,
  title={Qwen2-vl: Enhancing vision-language model's perception of the world at any resolution},
  author={Wang, Peng and Bai, Shuai and Tan, Sinan and Wang, Shijie and Fan, Zhihao and Bai, Jinze and Chen, Keqin and Liu, Xuejing and Wang, Jialin and Ge, Wenbin and others},
  journal={arXiv preprint arXiv:2409.12191},
  year={2024}
}

@inproceedings{wei2021contrastive,
  title={Contrastive learning for cold-start recommendation},
  author={Wei, Yinwei and Wang, Xiang and Li, Qi and Nie, Liqiang and Li, Yan and Li, Xuanping and Chua, Tat-Seng},
  booktitle={Proceedings of the 29th ACM international conference on multimedia},
  pages={5382--5390},
  year={2021}
}

@article{gai2024qarmquantitativealignment,
      title={QARM: Quantitative Alignment Multi-Modal Recommendation at Kuaishou}, 
      author={Gai, Kun and Yuan, Wei and Zhou, Guorui and Luo, Xinchen and Zhang, Xu and Zhang, Jiaqi and Huang, Rui and Cao, Jiangxia and Wang, Shiyao and Yu, Jinkai and Liu, Zhaojie and Lin, Hezheng and Zheng, Yichen and Sun, Tianyu and Hu, Qigen and Qiu, Changqing and Yan, Zhiheng and Zhang, Jingming and Zhang, Siming and Wen, Mingxing},
      journal={arXiv preprint arXiv:2411.11739},
      year={2024}
}

@article{deng2025onerec,
  title={Onerec: Unifying retrieve and rank with generative recommender and iterative preference alignment},
  author={Deng, Jiaxin and Wang, Shiyao and Cai, Kuo and Ren, Lejian and Hu, Qigen and Ding, Weifeng and Luo, Qiang and Zhou, Guorui},
  journal={arXiv preprint arXiv:2502.18965},
  year={2025}
}

@article{rajput2023recommender,
  title={Recommender systems with generative retrieval},
  author={Rajput, Shashank and Mehta, Nikhil and Singh, Anima and Hulikal Keshavan, Raghunandan and Vu, Trung and Heldt, Lukasz and Hong, Lichan and Tay, Yi and Tran, Vinh and Samost, Jonah and others},
  journal={Advances in Neural Information Processing Systems},
  volume={36},
  pages={10299--10315},
  year={2023}
}

@article{tan2025pcr,
  title={PCR-CA: Parallel Codebook Representations with Contrastive Alignment for Multiple-Category App Recommendation},
  author={Tan, Bin and Ge, Wangyao and Wang, Yidi and Liu, Xin and Burtoft, Jeff and Fan, Hao and Wang, Hui},
  journal={arXiv preprint arXiv:2508.18166},
  year={2025}
}

@article{bai2025chime,
  title={Chime: A compressive framework for holistic interest modeling},
  author={Bai, Yong and Xiang, Rui and Li, Kaiyuan and Tang, Yongxiang and Cheng, Yanhua and Liu, Xialong and Jiang, Peng and Gai, Kun},
  journal={arXiv preprint arXiv:2504.06780},
  year={2025}
}

@article{li2025bbqrec,
  title={Bbqrec: Behavior-bind quantization for multi-modal sequential recommendation},
  author={Li, Kaiyuan and Xiang, Rui and Bai, Yong and Tang, Yongxiang and Cheng, Yanhua and Liu, Xialong and Jiang, Peng and Gai, Kun},
  journal={arXiv preprint arXiv:2504.06636},
  year={2025}
}

@inproceedings{sheng2024enhancing,
  title={Enhancing Taobao Display Advertising with Multimodal Representations: Challenges, Approaches and Insights},
  author={Sheng, Xiang-Rong and Yang, Feifan and Gong, Litong and Wang, Biao and Chan, Zhangming and Zhang, Yujing and Cheng, Yueyao and Zhu, Yong-Nan and Ge, Tiezheng and Zhu, Han and others},
  booktitle={Proceedings of the 33rd ACM International Conference on Information and Knowledge Management},
  pages={4858--4865},
  year={2024}
}

@article{yan2025mim,
  title={MIM: Multi-modal Content Interest Modeling Paradigm for User Behavior Modeling},
  author={Yan, Bencheng and Chen, Si and Jia, Shichang and Liu, Jianyu and Liu, Yueran and Fu, Chenghan and Guan, Wanxian and Zhao, Hui and Zhang, Xiang and Zhang, Kai and others},
  journal={arXiv preprint arXiv:2502.00321},
  year={2025}
}

@inproceedings{zhang2026onetrans,
  title={Onetrans: Unified feature interaction and sequence modeling with one transformer in industrial recommender},
  author={Zhang, Zhaoqi and Pei, Haolei and Guo, Jun and Wang, Tianyu and Feng, Yufei and Sun, Hui and Liu, Shaowei and Sun, Aixin},
  booktitle={Proceedings of the ACM Web Conference 2026},
  pages={8162--8170},
  year={2026}
}

@inproceedings{zhu2025query,
  title={Query-LIFE: Query-aware Language Image Fusion Embedding for E-Commerce Relevance},
  author={Zhu, Hai and Guo, Yuankai and Dou, Ronggang and Liu, Kai},
  booktitle={Proceedings of the 31st International Conference on Computational Linguistics: Industry Track},
  pages={21--28},
  year={2025}
}

@article{yao2025saviorrec,
  title={Saviorrec: Semantic-behavior alignment for cold-start recommendation},
  author={Yao, Yining and Li, Ziwei and Xiao, Shuwen and Du, Boya and Zhu, Jialin and Zheng, Junjun and Kong, Xiangheng and Jiang, Yuning},
  journal={arXiv preprint arXiv:2508.01375},
  year={2025}
}

@article{zhang2025gpr,
  title={GPR: Towards a Generative Pre-trained One-Model Paradigm for Large-Scale Advertising Recommendation},
  author={Zhang, Jun and Li, Yi and Liu, Yue and Wang, Changping and Wang, Yuan and Xiong, Yuling and Liu, Xun and Wu, Haiyang and Li, Qian and Zhang, Enming and others},
  journal={arXiv preprint arXiv:2511.10138},
  year={2025}
}

@article{liu2025best,
  title={The Best of the Two Worlds: Harmonizing Semantic and Hash IDs for Sequential Recommendation},
  author={Liu, Ziwei and Wang, Yejing and Liu, Qidong and Zhang, Zijian and Chen, Chong and Huang, Wei and Zhao, Xiangyu},
  journal={arXiv preprint arXiv:2512.10388},
  year={2025}
}

@article{wan2026r3vae,
  title={R3-VAE: Reference Vector-Guided Rating Residual Quantization VAE for Generative Recommendation},
  author={Wan, Qiang and Yang, Ze and Yang, Dawei and Fan, Ying and Yan, Xin and Liu, Siyang},
  journal={arXiv preprint arXiv:2604.11440},
  year={2026}
}

@article{pan2026hisam,
  title={Hi-SAM: A Hierarchical Structure-Aware Multi-modal Framework for Large-Scale Recommendation},
  author={Pan, Pingjun and Zhou, Tingting and Lu, Peiyao and Fei, Tingting and Chen, Hongxiang and Luo, Chuanjiang},
  journal={arXiv preprint arXiv:2602.11799},
  year={2026}
}

@article{khrylchenko2026variable,
  title={Variable-Length Semantic IDs for Recommender Systems},
  author={Khrylchenko, Ivan and others},
  journal={arXiv preprint},
  year={2026}
}

@article{wu2025muse,
  title={MUSE: A Simple Yet Effective Multimodal Search-Based Framework for Lifelong User Interest Modeling},
  author={Wu, Bin and Yang, Feifan and Chan, Zhangming and Gu, Yu-Ran and Feng, Jiawei and Yi, Chao and Sheng, Xiang-Rong and Zhu, Han and Xu, Jian and Ye, Mang and Zheng, Bo},
  journal={arXiv preprint arXiv:2512.07216},
  year={2025}
}

@article{liu2026paletteid,
  title={PaletteID: Prototype-Composed Semantic Identifiers for Multimodal CTR Prediction},
  author={Liu, Huanyu and Chen, Baining and Liu, Hui and Li, Zengyang and Huang, Ziyi},
  journal={arXiv preprint arXiv:2607.29000},
  year={2026}
}

@article{zhu2025rankmixer,
  title={RankMixer: Scaling Up Ranking Models in Industrial Recommenders},
  author={Zhu, Jie and Fan, Zhifang and Zhu, Xiaoxie and Jiang, Yuchen and Wang, Hangyu and Han, Xintian and Ding, Haoran and Wang, Xinmin and Zhao, Wenlin and Gong, Zhen and others},
  journal={arXiv preprint arXiv:2507.15551},
  year={2025}
}

@article{hu2026quasid,
  title={Stop Treating Collisions Equally: Qualification-Aware Semantic ID Learning for Recommendation at Industrial Scale},
  author={Hu, Zheng and Chen, Yuxin and Pan, Yongsen and Yuan, Xu and Yin, Yuting and Wang, Daoyuan and Xia, Boyang and Luo, Zefei and Wang, Hongyang and Ni, Songhao and others},
  journal={arXiv preprint arXiv:2603.00632},
  year={2026}
}

@article{xu2025mmq,
  title={MMQ: Multimodal Mixture-of-Quantization Tokenization for Semantic ID Generation and User Behavioral Adaptation},
  author={Xu, Yang and others},
  journal={arXiv preprint arXiv:2508.15281},
  year={2025}
}

@article{liu2025discrec,
  title={DiscRec: Disentangled Semantic-Collaborative Modeling for Generative Recommendation},
  author={Liu, Chang and Bai, Yimeng and Zhao, Xiaoyan and Zhang, Yang and Feng, Fuli and Rong, Wenge},
  journal={arXiv preprint arXiv:2506.15576},
  year={2025}
}

\end{document}